\documentclass[journal, twoside]{IEEEtran} 

\usepackage[cmex10]{amsmath}
\usepackage{algorithmic}
\usepackage{url}

\usepackage[cmex10]{amsmath}
\usepackage{graphicx}
\usepackage{array}
\usepackage{amssymb}
\usepackage{subfigure}
\usepackage{verbatim}
\usepackage{epsfig}
\usepackage{cleveref}
\usepackage{color}
\usepackage{cite}

\IEEEoverridecommandlockouts                             

\begin{document}

\title{\textbf{\Large{Flexibility of Commercial Building HVAC Fan as Ancillary Service for Smart Grid}}}

\author{ \parbox{6 in}{\centering Mehdi Maasoumy$^\dagger$, Jorge Ortiz$^*$, David Culler$^*$ and Alberto Sangiovanni-Vincentelli$^*$
%
\thanks{ $^\dagger$Department of Mechanical Engineering, University of California, Berkeley, CA 94720-1740, USA. Corresponding author. Email: {\tt \small mehdi@me.berkeley.edu}}
\thanks{ $^{*}$Department of Electrical Engineering and Computer Sciences, University of California, Berkeley, CA 94720-1740, USA.}}
}


\markboth{Proceedings of Green Energy and Systems Conference 2013, November 25, Long Beach, CA, USA. }{This full text paper was peer reviewed at the direction of Green Energy and Systems Conference subject matter experts.}

\maketitle


\IEEEpeerreviewmaketitle

\begin{abstract}
In this paper, we model energy use in commercial buildings using empirical data captured through sMAP, a campus building data portal
at UC Berkeley.
We conduct at-scale experiments in a newly constructed building on campus.  By modulating the supply duct static pressure (SDSP) 
for the main supply air duct, we induce a response on the main supply fan and determine how much 
ancillary power flexibility can be provided by a typical commercial building.  We show that the consequent intermittent 
fluctuations in the air mass flow into the building does not influence the building climate in a human-noticeable way. 
We estimate that at least 4 GW of regulation reserve is readily available only through commercial buildings in the US. Based on predictions this value will reach to 5.6 GW in 2035.  
We also show how thermal slack can be leveraged to provide an ancillary service to deal with transient frequency fluctuations in the grid. 
We consider a simplified model of the grid power system with time varying demand and generation and present a simple control scheme to direct the ancillary service power flow from buildings to improve on the classical automatic generation control (AGC)-based approach. Simulation results are provided to 
show the effectiveness of the proposed methodology for enhancing grid frequency regulation.
\end{abstract}

\section{Introduction}
\label{sec:intro}
Total primary energy consumption in the world increased more than 27\% over the last decade; from 400 Quadrillion Btu in 2000 to 510 Quadrillion Btu in 2010~\cite{EIA}. A sustainable energy future requires significant and widespread penetration of renewable energy sources (RES) than the current level. However, the volatility, uncertainty, and intermittency of renewable energy sources present a daunting challenge to integrate them into the power grid at large scale.

Balancing generation and load instantaneously and continuously, given the randomness in the dynamics of generation and demand, is difficult. Minute-to-minute load variability results from random de/activation of millions of individual loads. Long-term variability results from predictable factors such as the daily and seasonal load patterns as well as more random events like shifting weather patterns. Generators also introduce unexpected fluctuations because they do not follow their generation schedules exactly and they trip unexpectedly due to a range of equipment failures~\cite{kirby2005frequency}.

Significant deviation in supply-demand balance can lead to large frequency deviation, which in turn jeopardizes the stability of the grid. To avoid this catastrophic event, several so called ``ancillary services" such as regulation and load following, have been formalized to better manage supply-demand balance at all time.
The Federal Energy Regulatory Commission (FERC) has defined such services as those ``necessary to support the transmission of electric power from seller to purchaser given the obligations of control areas and transmitting utilities within those control areas to maintain reliable operation of the interconnected transmission system.'' This quote highlights the importance of ancillary services for both bulk-power reliability and support of commercial transactions~\cite{kirby2005frequency}.

Buildings consume about 75\% of US electricity, with roughly equal shares for residential and commercial buildings~\cite{DOE-2010-Online}. Commercial buildings are suitable for providing ancillary services due to the following reasons: 1) More than 30\% of commercial buildings have adopted Building Energy Management System (BEMS) technology~\cite{EIA:Commercial} which facilitates the communication with the grid system operators for providing real-time ancillary services. The majority of these buildings are also equipped with variable frequency drives, which in coordination with BEMS, can manipulate the heating, ventilation and air conditioning (HVAC) system power consumption very frequently (in the order of several seconds). 2) Compared to typical residential buildings, commercial buildings have larger HVAC systems and therefore consume more electricity and present an opportunity for manipulating and controlling the building's power draw. HVAC system fans account for about 15\% of electricity consumed in commercial buildings. Since we can directly control their power-draw rate, upward or downward – fans are an ideal candidate for ancillary service.

\subsection{Related Works}
Model-based optimal control strategies such as Model Predictive Control (MPC) are promising for energy efficiency in buildings and for integrating time-of-use rates for shifting loads~\cite{maasoumy2011DSC, Maa:TotalAndPeak, MaasoumyHaghighi:EECS-2011-12,oldewurtel2010reducing,maasoumy2012optimal,6204325}. In a more recent paper~\cite{hao2013ancillary} simulation data show that commercial buildings can provide significant ancillary service for more robust operation of the grid. The same paper postulates 6.6 GW of regulation capacity from about 5 million commercial buildings in the US. In this paper, we attempt to verify this claim via experiments on a real building.

Regulation is a zero-energy service, making it an ideal candidate for supply by storage. In~\cite{kirby2005frequency}, storage technologies are acknowledged to be ideal suppliers of several ancillary services, including regulation. Storage using chemical batteries however, has two important drawbacks: 1) It is expensive and 2) it is not environmentally-friendly. There is an emerging consensus that flexible loads with thermal storage capabilities, also known as Thermostatically Controllable Loads (TCL) will play an important role in regulating grid frequency and in effect, enable deep penetration of RES.


Control strategies for aggregated heterogeneous thermostatically-controlled loads (TCL) for ancillary service is discussed in~\cite{koch2011modeling,mathieu2012state}. TCLs, such as refrigerators, air conditioners, and electric water heaters, are well-suited to direct load control (DLC) and demand-response (DR) programs that require loads to both decrease and increase power consumption because they are capable of storing thermal energy, much like a battery stores chemical energy. Fully responsive load control is discussed in~\cite{callaway2011achieving} in the context of TCLs and plug-in electric vehicles. Despite several challenges of using loads for system services, several key advantages include: $1)$ Less variability associated to a very large number of small loads when compared to that of a small number of large generators~\cite{kirby2003spinning}; $2)$ Instantaneous response of loads to operator requests, versus slow response of generators to significant output changes~\cite{kirby2003spinning}; $3)$ Reducing overall grid emissions by using loads to provide system services~\cite{strbac2008demand}. 

This paper focuses on an important ancillary service called, \emph{frequency regulation}, which is provided by power sources online, on automatic generation control (AGC), that can respond rapidly to system-operator requests for up and down movements. Regulation is a zero-energy service that compensates for minute-to-minute fluctuations in load and corrects for unintended fluctuations in generator output to comply with control performance standards of the North-American Electric Reliability Corporation (NERC)~\cite{nerc}. 

\subsection{Our Contributions}
In this paper, we analyze the energy use in a commercial building located on the UC Berkeley campus, and ground our assumptions on empirical data. We show through at-scale experiments that \textit{regulation services}, with a response time of a few seconds and duration of several minutes (to hours), can be provided using the supply fans of building HVAC systems, free of charge, i.e. without an increase in total energy consumption and without causing sensible temperature change in buildings.
We then consider a simplified model of the power system with \textit{uncertain} demand and generation. We demonstrate, through simulation, how ancillary services can improve grid frequency regulation.

\section{At-Scale Experiments and Building Data Analysis}\label{sec:buildanalysis2}
We examine the dynamics of the building from a power-draw perspective and look specifically at the effects of supply-fan
speed modulation on the internal temperature of rooms in the building. The supply fans to the building are large
and hence draw significant power. Therefore, they provide a point of control for the response of a building (or many buildings) to
an ancillary-service signal coming from the grid.  The main challenges include 1) keeping the pressure in the air ducts within
a specific safety boundary and 2) minimize changes to internal temperature that may make occupants uncomfortable.

In order to address these constraints we run a number of experiments where we vary the speed of the fan at different rates,
for different lengths of time.  We show that supply-fan speed modulation, over short
time periods, has little affect on the comfort of the building occupants.  In fact, the effects are not observable
on the internal temperature of the spaces.  We also show that the supply fan responds very quickly to a change in pressure set-point
changes, indicating that this 
is a viable option for providing grid-level ancillary response services. In this paper we focus on very fast (minute-by-minute) manipulation of fan speed. It is shown in~\cite{hao2013ancillary} that 15\% of fan power modulation imposes less than 0.2 temperature deviation on the room air temperature. However, due to the complexity and scale of commercial buildings HVAC systems, the simplified and \emph{lumped} building models may fail to
capture how fast fan-speed changes effect the building climate.  Hence, we run experiments to corroborate the claim that fast modulation of the fan speed
does not affect room temperatures in a human-sensible way.

We run our experiments in a building constructed four years ago (2009) on the UC Berkeley Campus named Sutardja Dai Hall (SDH).
It is a 141,000 square foot modern building that houses several laboratories, including a nano-fabrication lab, dozens of classrooms and
collaborative work spaces.  It contains a Siemens building management system called Apogee~\cite{apogee} which is connected to an
sMAP server~\cite{smap} co-located with the Apogee server.  We run our experiments using a control platform built on sMAP, whereby control
points are set programmatically upon approval from the building manager.

\subsection{Main Building Air Supply Fan}

\begin{figure}[t!] 
\centering
\includegraphics[width=\columnwidth]{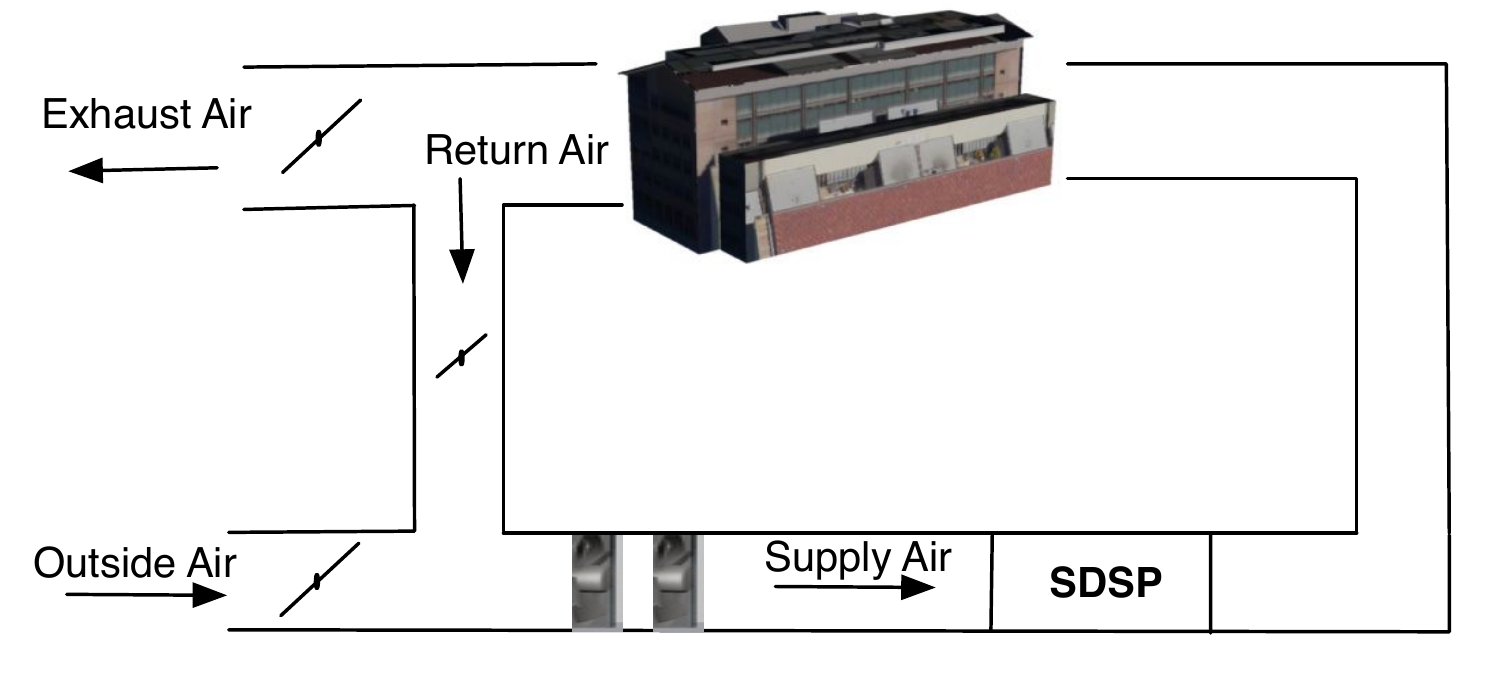}
\caption{Schematic of closed-loop air circulation in the ducts of the test-bed in hand, Sutardja-Dai Hall on the campus of University of California, Berkeley. We changed the supply duct static pressure (SDSP) set-point to change the fan speed, and consequently fan power draw.}
\label{fig:fans_feed_building}
\end{figure}

The main supply fans that feed SDH can spin at variable speeds, directly proportional to the maximum rated
power draw --  67 KW or about 7\% of the total power consumed in that building, per fan.  Moreover,
we can directly control its power-draw rate, positive or negative -- making it an ideal candidate for
ancillary service.  Figure~\ref{fig:fans_feed_building} shows a high-level schematic of the
two fans that feed Sutardja-Dai Hall at UC Berkeley.  The speed of the fans is indirectly controlled by setting the
supply-duct static pressure (SDSP) set point value.

The main high-level challenges for using supply fans are related to safety and comfort. Because of safety considerations, we use the supply-duct pressure set point to control fan speed.  The safety margin for the pressure set point ranges from 1.2-1.9 (Inch of Water). By changing the SDSP set point, the fan speed either increases or decreases to adhere to the specified pressure value. We also test how quickly the fans respond to changes to the pressure set point and model the response in section~\ref{sec:fan_model}.



To test the effects on internal climate we use the internal temperature of the rooms in the building as a proxy for this measure.  Typically, the internal temperature varies between 68-72 $^oF$.  We run three experiments where we change the pressure setpoint at different intervals. The first experiment changes the setpoint between 1.2 and 1.9 every minute for 15 minutes.  The second experiment changes the set point every 2 minutes for about half an hour.  The third changes the set point every 3 minutes, also for half an hour.  The first experiment (exp1) took place on June 6th, 2013, between 3:00 PM - 3:15 PM, the second and third experiments (exp2), and (exp3) took place on June 7th, 2013, from 11:01 AM - 11:25 AM, and from 2:55 PM - 3:55 PM, respectively. We deliberately ran experiments during different times of the day in order to observe the effects of changes in the SDSP on the internal temperature of the rooms.

\begin{figure}[t!] 
\centering
\includegraphics[width=1.0\columnwidth]{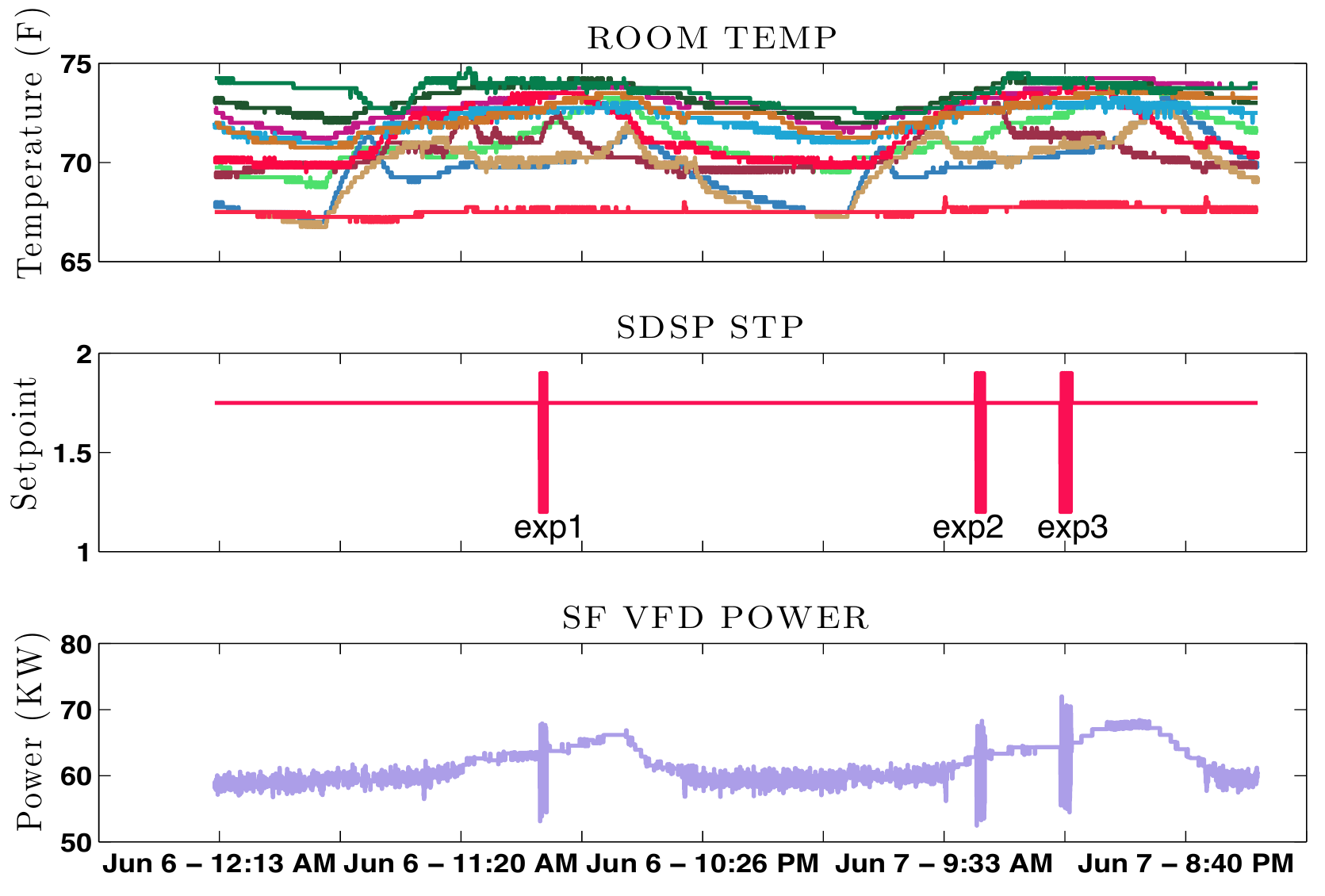}
\caption{Three experiments were performed on June 6 and 7, 2013. The middle figure shows the SDSP set-point (control input), the bottom figure shows the corresponding power draw of HVAC fan, and the top figure shows the temperature of 10 randomly-sampled rooms in the building.}
\label{fig:jun7_sdsp_power_temps}
\end{figure}

Figure~\ref{fig:jun7_sdsp_power_temps} shows the results of the experiments.  The top graphs shows the temperature of 10 random rooms in
the building.  The center graph shows the SDSP set point.  Notice the values fluctuating when the experiment is run.  The third graph
shows the power drawn by the supply fan, in response to changes in the setpoint.  Notice, the temperature is not noticeably affected
by our experiments and always stays within the comfort range.  These experiments confirm our hypothesis and support the use
of the supply fans for responding to grid activity.  In aggregate, we can control almost 14\% of the total energy consumed in this building.
Across a stock of buildings, this presents a significant amount of shiftable load.




\subsection{Total Ancillary Service Available in the US}
Figure~\ref{fig:jun7_sdsp_power_temps} also shows that in essence, each fan can provide about 18\% of its nominal power draw as flexibility. This equals 12 kW out of 67 kW power draw for each fan, and in total 24 kW for the whole SDH building. Change of this power draw can be done very fast, hence we have for all $t$
\begin{equation}
\max ~~ |P_{\mathrm{fan}}(t)-P_{\mathrm{fan}}(t- \delta)| = 24 ~kW
\end{equation}
where $\delta$ depends on various factors such as the HVAC fan size and mechanical inertia. For the SDH building we observed $\delta \simeq 1 sec$.
 
According to the latest survey on energy consumption of commercial buildings, performed in 2003~\cite{EIA:Commercial}, there are 4.9 million commercial buildings in the US which cover a total area of about 72 billion square foot. Almost 30\% of these buildings are equipped with variable frequency drive fans. Assuming the same fan power consumption flexibility per square foot to that of SDH for all commercial buildings, we estimate that at least 4 GW of fast ancillary service is readily available in the US at almost no cost, based on the 2003 data. Commercial building floor space is expected to reach 103 billion sq. ft. in 2035~\cite{EIA:2012}. With the same assumption of the above calculations, about 5.6 GW of regulation reserve will be available in 2035.



\subsection{Demand-Response Load Shedding}

\begin{figure}[t!] 
\centering
\includegraphics[width=1.0\columnwidth]{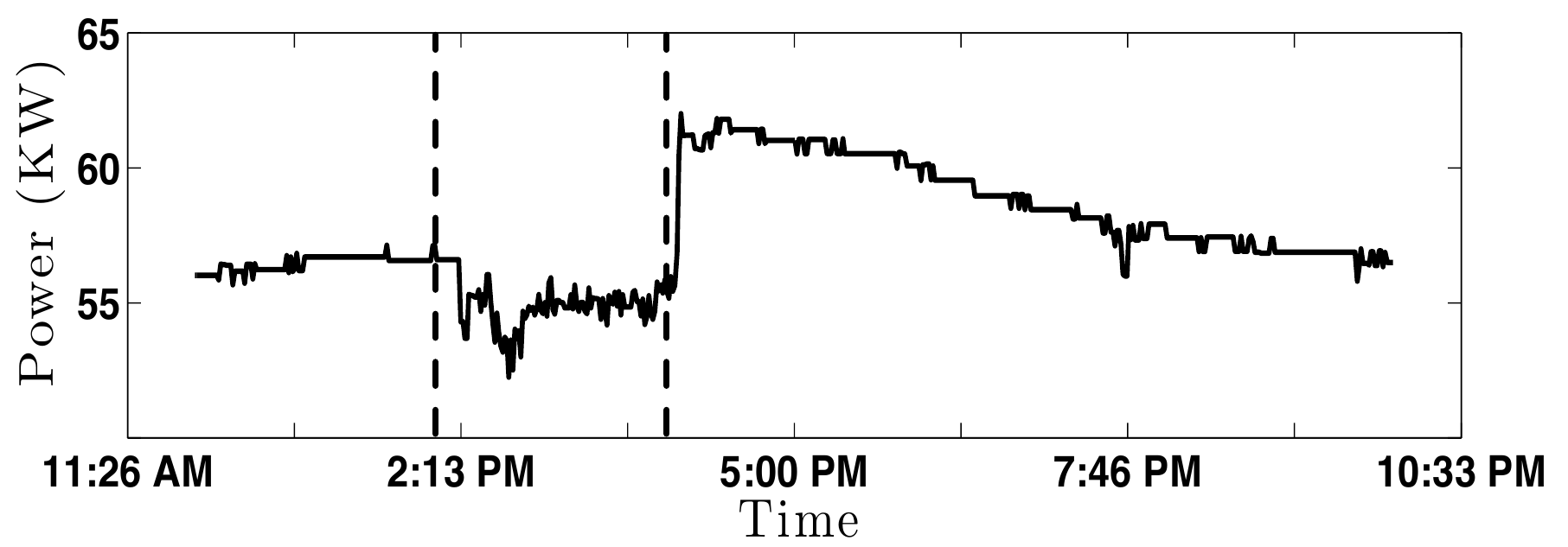}
\caption{Supply fan power draw during a demand-response event on September 11, 2012.  The region between the dashed
lines indicates the demand-response time frame.  A reduction in building activity induces a nearly 10\% reduction in
the power draw of the supply fan.}
\label{fig:DR_event1}
\end{figure}

Demand-response (DR) offers a way to shed load by inducing building occupants to actively reduce their energy consumption over some
period of time.  Typically the event is known in advance and the building manager announces the start and end of the event
by broadcasting a power-reduction event to the occupant -- pleading participation in the event.  Such an event is administered
several times a year in Sutardja Dai Hall.  Figure~\ref{fig:DR_event1} shows the effect of a DR event on the power-draw of the supply
fan during a DR event administered on September 11, 2012 from 2PM-4PM.

The average-power draw of the fan during normal operation is about 66 KW and dropped to 59 KW during the demand-response event --
a decreases of 9.7\%.  The average drop is about 6.5 KW for two hours, or about 0.68\% of the total average consumption of the building.
The total power reduction in the building, due to the event, is about 30\%.  The supply fan only contributes a small fraction of the drop
and is not a directly
targeted during the event, only a reduction of end-use devices is targeted.  For example, this event called for lights being turned off,
printers and coffee machines being powered down and laptops to run on batteries instead of outlet power.

As previously mentioned, the supply fans represent 14\% of the total power draw of the building and we have shown that a reduction in fan speed does not affect the occupants, if reduced with an almost zero-mean modulation. Our experiments demonstrate the utility of controlling the supply fan, both for DR events \emph{and} for more immediate responses to grid events.  They require no coordination with the occupants, no a-priori knowledge of an upcoming event, and does not affect the operation of the building. It also represents nearly half the energy reduction obtained through a DR event. We conclude that supply-fan control for load shedding in either event is both useful and prudent.

\section{Mathematical Model of HVAC Fan}
\label{sec:fan_model}
As explained in previous section, the experiments on the fan was performed by changing the pressure set point of the supply duct static pressure (SDSP). In order to track the SDSP signal the two fans change their speeds and consequently their power consumption changes. In this section we identify a single-input single-output (SISO) model of the fan with SDSP as input and fan power consumption as output. We consider an ARX model and use the historical data obtained from the experiments on the SDH building to identify the parameters of the fan model. 

Consider an ARX model with orders of model $\lbrace n_a, n_b, n_k\rbrace$, given by
\begin{align}\label{eq:ARX}
y(t) + &a_1y(t-1)+...+a_{n_a}y(t-n_a)=\\
&b_1u(t-1)+...+b_{n_b}u(t-n_k-n_b+1)+e(t) \notag
\end{align}
where $y(t) \in \mathbb{R}$ is the output, and $u(t) \in \mathbb{R}$ is the input, and $e(t)$ is the white noise at time $t$. $a_1,...,a_n$ and $b_1,...,b_n$ are the parameters to be estimated. $n_a$ is the number of poles of the system, $n_b-1$ is the number of zeros of the system, and $n_k$ is the delay value. \eqref{eq:ARX} can be written compactly as
\begin{equation}
y(t)=\phi^{T}(t)\theta
\end{equation}
where $\phi(t)$ and $\theta$ are given by
\begin{equation}
\phi(t)=\begin{bmatrix}
-y(t-1)\\ \vdots \\ -y(t-n_a) \\ u(t-n_k-1) \\ \vdots \\ u(t-n_k-n_b)
\end{bmatrix} ~~~~~ \theta=\begin{bmatrix}
a_1 \\ \vdots \\ a_n \\ b_n \\ \vdots \\b_m
\end{bmatrix} 
\end{equation}
\noindent where $\phi(t) \in \mathbb{R}^{n_a+n_b}$ is the data vector comprised of augmented vector of $n_a$ output and $n_b$ input measurements also known as \textit{regressors} vector and $\theta \in \mathbb{R}^{n_a+n_b}$ is the parameter vector. The linear ARX model thus predicts the current output $y(t)$ as a weighted sum of its regressors. The identification process solves for vector $\theta$ using M observations.~\eqref{eq:ARX} for M observations can be written as:
\begin{equation}
\mathbf{Y}(t)=\mathbf{\Phi}^{T}(t)\theta
\end{equation}
in which augmented matrices $\mathbf{Y}$ and $\mathbf{\Phi}$ are given by:
\begin{align}
\mathbf{\Phi}&=[\phi(1)~ \phi(2)~ \ldots ~ \phi(M)]\\
\mathbf{Y}&=[y(1)~ y(2)~ \ldots ~ y(M)]^T
\end{align}
 
In the case of no noise, $\theta$ can be solved for using $M\geq n_a + n_b$ observations of the system, with the assumption of persistent excitation~\cite{ljung1999system}.

\begin{figure}[t] 
\centering
\includegraphics[width=1.0\columnwidth]{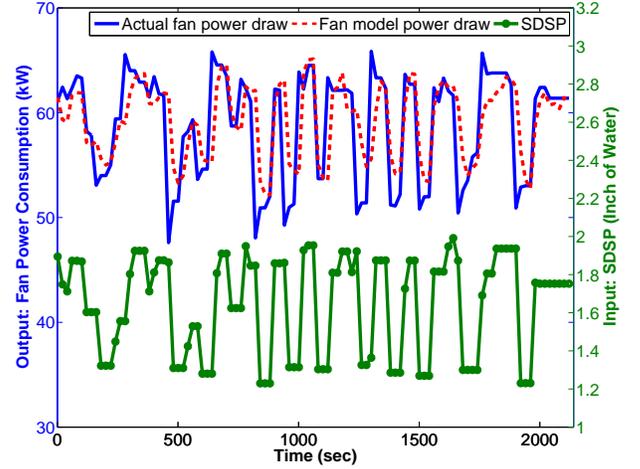} 
\caption{Supply duct static pressure (SDSP) is varied within [1.2,1.9] as an input to the system leading to the variation in the output i.e fan power consumption. Data collected from this experiment is used to identify the SISO model of the fan.}
\label{fig:ARX}
\end{figure}
\subsection{Identification Results}
We perform the identification process on the introduced ARX model and historical data obtained from the experiments detailed earlier. The input signal (SDSP) and the measurements as well as the obtained fan model output are shown in Figure~\ref{fig:ARX}

The transfer function obtained here can be used to track the ancillary signal from system operator by manipulating the HVAC fan SDSP set-point and consequently through the fan VAV system, the instantaneous power draw of the fan.

\section{Power System Characteristics}\label{sec:PowerSys}

\subsection{Mathematical Models of Power System Components} \label{sec:MathModel}
\begin{figure}[t] 
\centering
\includegraphics[width=1.0\columnwidth]{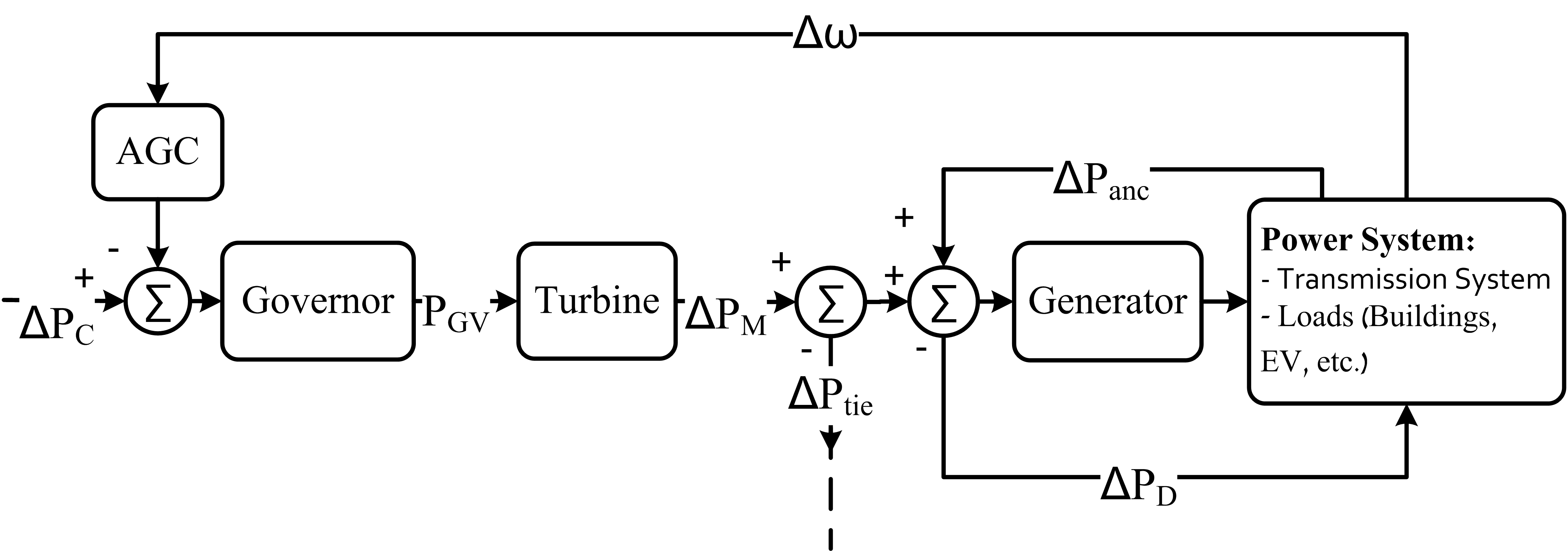}
\caption{Block diagram of power system and its relation to governor, turbine, generator and the AGC for each control area.}
\label{fig:BlockDiag}
\end{figure}

In this paper we use the detailed models of power system components that have been developed in the literature~\cite{debs1988modern,vittal1999power,kundur1994power}. Here we present a detailed governor, turbine and generator model as shown in Fig.~\ref{fig:BlockDiag}. Basic variables of a power system which are also used in the formulations are reported in Table~\ref{Table:var}.
\begin{table} [t]
\centering
\caption{Basic Power Systems Nomenclature}
  \begin{tabular}{lp{6.1cm}} 
    \hline
    Variables & Description \\ \hline
    $P_M$ & Mechanical power input  \\ 
    $P_M^o$ & Desired real power generation \\
		$P_G + j Q_G$ & Generated electric power with real (P) and reactive (Q) parts \\
		$\Delta P_G$ & Increase in demand (at rated generator MVA) \\
		$P_D$ & Load \\
		$\Delta P_D$ & Input disturbance due to load changes \\
		$\Delta P_C$ & Speed changer position feedback control signal\\
		$\omega$ & Angular speed \\
		$\omega_{des}$ & Rated (desired) frequency \\
    \hline
  \end{tabular}
	\label{Table:var}
\end{table}
$\Delta P_C$ is a control input which acts against increase or decrease in the power demand to regulate the system frequency. $\Delta P_D$ denotes the fluctuations in the power demand which is considered here as an exogenous input (disturbance). 

\subsection{Governor Model}
A generic model of a governor contains three time constants. The overall transfer function is given by
\begin{equation}
T_{Gov}(s)=\frac{(1+sT_{2})}{(1+sT_{1})(1+sT_{3})}
\end{equation}

Usually, mechanical-hydraulic governors have $T_{2}=0$ with typical values of $T_1\in[0.2,0.3]$ and $T_3=0.1$. Electro-hydraulic governors without steam feedback have typical time constants as follows: $T_1=T_2=0$ and $T_3\in[0.025,0.1]$. Electro hydraulic governors with steam feedback utilize a feed-forward mechanism (and hence the numerator term with $T_2$). For these one obtains typical values as follows: $T_1=2.8$. $T_2=1.0$, and $T_3=0.15$~\cite{debs1988modern}.

\subsection{Turbine Model}
Turbines are grouped into steam and hydro turbines. The transfer function model for turbines is given by
\begin{equation}
\frac{\Delta P_M}{P_{GV}} = K_1F_1 + K_2F_1F_2 + K_3F_1F_2F_3 + K_4F_1F_2F_3F_4
\end{equation}
\noindent where $F_1,F_2,F_3,F_4$ are transfer functions corresponding to steam chest, piping system, re-heaters and cross-over mechanisms, respectively, and are given by
\begin{align}
F_1(s)&=\frac{1}{1+sT_4}  &F_2(s)=\frac{1}{1+sT_5}\\
F_3(s)&=\frac{1}{1+sT_6}  &F_4(s)=\frac{1}{1+sT_7}
\end{align}
The basic time constant associated with steam turbines is $T_4$ which corresponds to that of the \textit{steam chest}. For non-reheat steam turbines, this is the only time constant needed. The time constants $T_5$, $T_6$, and $T_7$, are associated with time delays of piping systems for re-heaters and cross-over mechanisms. Coefficients $K_1$, $K_2$, $K_3$, and $K_4$ represent fractions of total mechanical power outputs associated with \emph{very high}, \emph{high}, \emph{intermediate}, and \emph{low} pressure components, respectively~\cite{debs1988modern}. 
\subsection{Generator Model}
The dynamics of the generator is given by the following transfer function
\begin{equation}
F_{Gen}=\frac{1}{D+sM}
\end{equation}
where constants $D$ and $M$ represent the damping coefficient and the inertia of governor, respectively.

\subsection{Speed Regulation}
Ramping up or down of turbines has to be controlled and slowly performed with a governor to prevent damage. Uncontrolled acceleration of the turbine rotor can lead to an overspeed trip which can cause unintended closure of the nozzle valves that control the flow of steam to the turbine. In the event of such failure, the turbine may continue accelerating and eventually break apart. For stable operation of the electrical grid of North America, power plants operate with a five percent \emph{droop speed control} during normal and synchronized operation. This means the full-load speed is 100\% and the no-load speed is 105\%. Normally the changes in speed are minor due to inertia of the total rotating mass of all generators and motors. Power output adjustments are made by slowly raising the droop curve which is done, by increasing the spring pressure on a centrifugal governor~\cite{whitaker2002ac}.

\subsection{Inherent Ramping Inertia of Generating Units}
Power system includes generation units. Traditional coal-based power plants have a slow response. However, hydro units on the other hand have faster response. Inherent variability of load and generation which has become more significant with the introduction of RES requires some level of flexibility in the power system to fill the gap between generation and consumption. Generating units however, due to the inertia of mechanical components have a high response time. It is shown in~\cite{wang1993effects} that incorporating ramping-rate characteristics into the AGC design problem, can enhance the overall performance of the system. While~\cite{wang1993effects} centers on unit commitment and dispatch processes, we in this paper focus on fast frequency regulation services. 

%

%

\section{Frequency Regulation with Ancillary Service}\label{sec:Pcontrol}
\begin{figure}[t] 
\centering
\includegraphics[width=1.0\columnwidth]{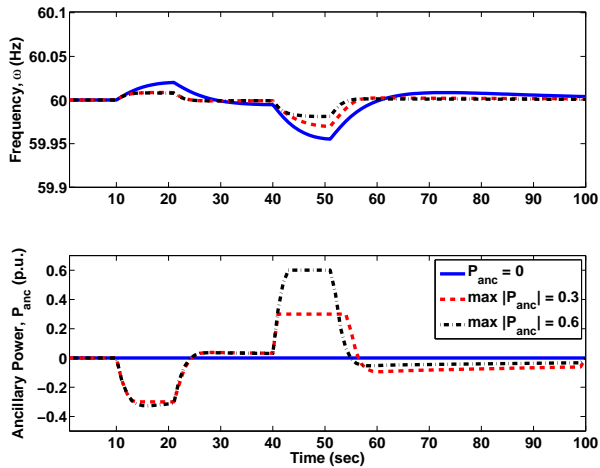} 
\caption{Frequency regulation shown for three scenarios in response to a change in demand. $\Delta P_D=0.5 (p.u.)$ and $\Delta P_D=-1 (p.u.)$ are introduced to the system for time T=[10,20], and T=[10,20], respectively.}
\label{fig:FreqReg}
\end{figure}

To demonstrate the impact of ancillary service on frequency regulation, we perform the following set of simulations. We consider one control area and use a simple proportional control to determine the flow of ancillary service. 
\begin{equation}
\hat{P}_{\mathrm{anc}}(\omega) =
  \begin{cases}
   -K_p .(\omega - \omega_{\mathrm{des}}) & \text{if } \omega \geq \omega_{\mathrm{des}} \\
   -K_p .(\omega - \omega_{\mathrm{des}}) & \text{if } \omega < \omega_{\mathrm{des}}
  \end{cases}
\end{equation}
Since the maximum ancillary power is limited by the available aggregate from buildings, we address the saturation by
 
\begin{equation}
P_{\mathrm{anc}}(\omega) =
  \begin{cases}
   \max \lbrace \hat{P}_{\mathrm{anc}}, \underline{P_{\mathrm{anc}}} \rbrace & \text{if } \omega \geq \omega_{\mathrm{des}} \\
   \min \lbrace \hat{P}_{\mathrm{anc}}, \overline{P_{\mathrm{anc}}} \rbrace & \text{if } \omega < \omega_{\mathrm{des}}
  \end{cases}
\end{equation}
\noindent where $\underline{P_{\mathrm{anc}}}\leq0$, and $\overline{P_{\mathrm{anc}}}\geq0$ are the minimum and maximum available ancillary power. Figure~\ref{fig:FreqReg} shows the result of simulations for three cases: 1) without ancillary power, 2) with ancillary power $\overline{P_{\mathrm{anc}}}=-\underline{P_{\mathrm{anc}}}=0.3 (p.u.)$ and 3) with ancillary $\overline{P_{\mathrm{anc}}}=-\underline{P_{\mathrm{anc}}}=0.6 (p.u.)$. Note that the values of 0.3 (p.u.) and 0.6 (p.u.) are considered for comparison purposes to show the effect of increasing the ramping rate of ancillary power on the control performance. We consider the following constant parameters for the model inertia $M=132.6 ~(MW.sec)$, damping $D=0.0265 ~(p.u.)$, $T_1=0.1$, $T_2=0$, $T_3=0.1$, $T_4=1$, $K_1=1$, $R=0.05$, and $K_p=45$. As observed from Figure~\ref{fig:FreqReg}, the performance of system improves by introduction of ancillary service, and by increasing the value of available ancillary power.

\section{Conclusion And Future Work} \label{sec:Conclusion}

We presented an empirical analysis on ancillary service potential which can be provided by a typical commercial building. The results of the analysis were used in a proportional control scheme in addition to the conventional AGC. We showed using simulations how ancillary service with even a simple control strategy improves the performance of the power systems by controlling the flow of ancillary services from buildings.

In the future we intend to develop a model predictive control framework that computes the optimum flow of ancillary service taking into account the mathematical model of the power system, the real-time frequency of the system, the inherent (slow) ramping rate of generation units, the fast ramping rate of ancillary service from buildings (as identified in this paper), maximum amount of available ancillary service, load forecasting, and energy pricing. Short-term load forecasting can be advantageous for designing more efficiently AGC, for optimal power flow, for security assessment, for unit commitment, for HTC, and for load/energy management. We are also developing an algorithm to obtain the theoretical maximum flexibility of building HVAC systems for longer time spans at any time, in an online fashion, given information such as outside climate, building occupancy schedules, current indoor temperature of the building, and the price of energy as well as reward from utility to the building for providing such flexibility. We are also working on developing a contractual framework that could be used by the building operator and the utility to declare flexibility on one side and reward structure on the other side.

\section{Acknowledgements}
This work was supported by the National Science Foundation under grants CPS-0931843 (ActionWebs), and CPS-1239552 (SDB). Mehdi Maasoumy is funded by the Republic of Singapore's National Research Foundation through a grant to the Berkeley Education Alliance for Research in Singapore (BEARS) for the Singapore-Berkeley Building Efficiency and Sustainability in the Tropics (SinBerBEST) Program. BEARS has been established by the University of California, Berkeley as a center for intellectual excellence in research and education in Singapore. Alberto Sangiovanni Vincentelli is supported in part by the TerraSwarm Research Center,one of six centers administered by the STARnet phase of the Focus Center Research Program (FCRP),a Semiconductor Research Corporation program sponsored by MARCO and DARPA. 

%


\bibliographystyle{IEEEtran} 
\bibliography{paperbib}

\vspace{-14 mm}
\begin{IEEEbiography}[{\includegraphics[width=25mm,height=32mm,keepaspectratio]{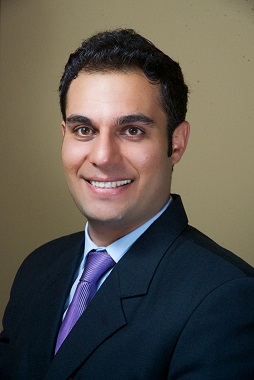}}]
{Mehdi Maasoumy} is a PhD Candidate at the Mechanical Engineering Department of University of California at Berkeley. His research interests include modeling and model-based optimal control of linear, nonlinear and hybrid systems, with applications in Energy Efficient Building Control Systems, Smart Grid and Aircraft Electric Power Distribution System. He has a B.S. degree in 2008 from Sharif University of Technology in Iran and a M.S. degree in 2010 from University of California at Berkeley both in Mechanical Engineering. He is a student member of IEEE and ASME. He is the recipient of the Best Student Paper Award at the International Conference on Cyber-Physical Systems (ICCPS) 2013, and the Best Student Paper Award Finalist at the ASME Dynamics and Control System Conference (DSCC) 2013.
\end{IEEEbiography} 
\vspace{-15 mm}

\begin{IEEEbiography}[{\includegraphics[width=25mm,height=32mm,clip,keepaspectratio]{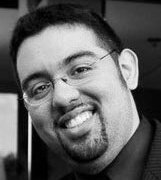}}]
{Jorge Ortiz} is a PhD Candidate in the Computer Science division at the University of California at Berkeley, working
with Professor David Culler on operating systems and networking.  He has an M.S. in Computer Science from UC Berkeley 
and a B.S. in  Electrical Engineering and Computer Science from M.I.T.  His research interests include energy-efficient buildings,
sensor networks, and cyber-physical systems.  He spent several years at Oracle and IBM prior to his time at Berkeley and 
worked on nano-satellite software at Nanosatisfi while at studying at Berkeley. Jorge is a member of the Software 
Defined Buildings (SDB) group UC Berkeley and will be a research scientist at IBM Research at the end of 2013.
\end{IEEEbiography} 
\vspace{-13 mm}

\begin{IEEEbiography}[{\includegraphics[width=25mm,height=32mm,clip,keepaspectratio]{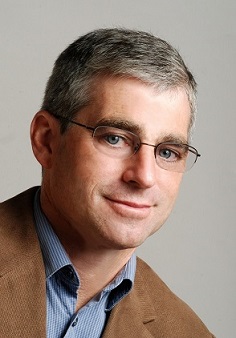}}]
{David Culler} is Chair of Electrical Engineering and Computer Sciences, and Faculty Director of i4energy at the University of California, Berkeley.  Professor Culler received his B.A. from U.C. Berkeley in 1980, and M.S. and Ph.D. from MIT in 1985 and 1989.   He was the Principal Investigator of the DARPA Network Embedded Systems Technology project that created the open platform for wireless sensor networks based on TinyOS,  co-founder and CTO of Arch Rock Corporation and the founding Director of Intel Research, Berkeley.  He is currently focused on utilizing information technology to address the energy problem and is co-PI on the NSF CyberPhysical Systems projects LoCal and ActionWebs and PI on Software Defined Buildings.
\end{IEEEbiography} 

\vspace{-123 mm}

\begin{IEEEbiography}[{\includegraphics[width=25mm,height=32mm,clip,keepaspectratio]{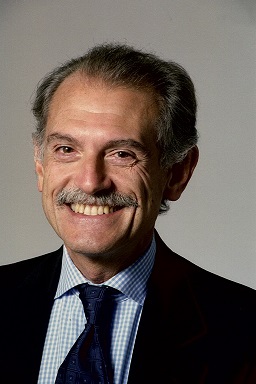}}]
{Alberto Sangiovanni Vincentelli}, the Buttner Chair, UC Berkeley, IEEE Fellow, Member of the National Academy of Engineering, co-founder of Cadence and Synopsys, the two largest EDA companies, is a member of the Board of Directors of Cadence, Sonics, ExpertSystems, and KPITCummins, of the Science and Technology Advisory Board of GM, of the Technology Advisory Council of UTC, of the Executive Committee of the Italian Institute of Technology, of the Scientific Council of the Italian National Science Foundation and President of the Group of 7 Research Supervisors of the Italian Government. He received the Kaufman Award for "pioneering contributions to EDA" and the IEEE/RSE Maxwell Medal "for groundbreaking contributions that had an exceptional impact on the development of electronics and electrical engineering or related fields".

\end{IEEEbiography}

\end{document}